# Improving VANET Protocols via Network Science

Romeu Monteiro[1,2], Susana Sargento[1]
[1]Instituto de Telecomunicações - Universidade de Aveiro
Aveiro, Portugal
{romeumonteiro7, susana}@ua.pt

Wantanee Viriyasitavat[2], Ozan K. Tonguz[2]
[2]ECE Department - Carnegie Mellon University
Pittsburgh, PA, USA
{wviriyas, tonguz}@ece.cmu.edu

*Abstract*— **Developing routing protocols for Vehicular Ad Hoc Networks (VANETs) is a significant challenge in these large, self-organized and distributed networks. We address this challenge by studying VANETs from a network science perspective to develop solutions that act locally but influence the network performance globally. More specifically, we look at snapshots from highway and urban VANETs of different sizes and vehicle densities, and study parameters such as the node degree distribution, the clustering coefficient and the average shortest path length, in order to better understand the networks' structure and compare it to structures commonly found in large real world networks such as small-world and scale-free networks. We then show how to use this information to improve existing VANET protocols. As an illustrative example, it is shown that, by adding new mechanisms that make use of this information, the overhead of the urban vehicular broadcasting (UV-CAST) protocol can be reduced substantially with no significant performance degradation.**

## I. Introduction

Developing suitable protocols for message dissemination in Vehicular Ad Hoc Networks (VANETs) is a challenging problem. Recent research on large scale self-organized distributed networks in nature (e.g., self-organizing biological networks) and social networks has shed new light on how these networks communicate in efficient ways in addition to paving the way for the development of a new perspective - commonly known as network science - for the study of these networks. Some of the most relevant studies include the celebrated work of Watts & Strotgatz [1] on small-world networks, and the well known work of Albert & Barabási [2] exploring scale-free networks. Kleinberg's work [3], building upon these earlier works, has shown how the structure of small-world networks relates to the optimal routing protocol.

Our goal in this paper is, by considering snapshots in time of urban and highway VANETs, to study the graphs they form from a network science perspective, analyzing parameters such as the nodes' degree distribution, the clustering coefficient, and the average shortest path length. The ultimate goal is to get useful information from these parameters to compare these networks to small-world and scale-free networks, and to use that information to improve the existing communications protocols for VANETs. This way, we will have a detailed description of the fundamental graph properties of urban and highway VANETs. We will show that this information will be instrumental for enhancing the performance of network protocols in vehicular networks.

In this paper, we analyze real data on the aforementioned parameters as a function of the vehicle density and the network area or length. This allows integration of local information such as the degree distribution of the nodes, and global information such as vehicle density and corresponding connectivity. Consequently, one can design adaptive protocols which work in a distributed way with local information, but that influence the performance of the network on a global level. We test this approach in the following way: first, we get the information regarding the clustering coefficient of urban networks and the relationship between node degree and connectivity; then, we use this information to improve the performance of the recently developed urban vehicular broadcast protocol (UV-CAST) [4] in extreme regimes of very low or very high vehicle density; finally, we evaluate this approach and show that this approach provides a very significant reduction on the overhead in the network with no performance degradation. We also analyze VANETs from a network science perspective and show that urban and highway VANETs are not scale-free networks since their node degree distributions can be modeled by Gaussian distributions. We also show that, even though there seems to be a possibility for a small-world behavior for low density VANETs, the connectivity is too low to profit from that behavior.

The remainder of the paper is organized as follows. Section II introduces the main differences between urban and highway VANETs and summarizes the information on the data used to emulate real world data. Section III analyzes several network science parameters for the two scenarios, accompanied by comparisons with theoretical models. Section IV elaborates on possible uses for the data and models analyzed in section III and applies them to improve the UV-CAST [4] protocol. Section V presents concluding remarks and highlights the possible areas for future work.

## II. Urban vs Highway VANETs

When studying the properties of VANETs one has to consider the two main scenarios - urban and highway - and how they impact the communication process. The main differences include: 1) Dimensionality (D), (i.e., 1D vs. 2D topology) has consequences on the number of vehicles within 1-hop distance, packet routing and vehicle clustering; 2) Signal Propagation, which is mostly line-of-sight on highways, but can include reflections and shadowing in urban environments; 3) Traffic patterns, influenced by factors such as the speed of the vehicles, the stability of their neighborhoods, and traffic congestion.

This work was funded in part by the Portuguese Foundation for Science and Technology under the Carnegie Mellon | Portugal program (grant SFRH/BD/51633/2011)

## A. Empirical data

In this paper we collect and analyze sets of data for both urban and highway VANETs in order to analyze the graph representation of these networks. These data sets contain the positions of the vehicles in different time moments, representing real life traffic data. In the case of urban VANETs, we use data generated by the Cellular Automata model (CA model) presented in [5], which uses a Manhattan grid model and emulates real vehicle patterns of movement in urban areas. This model includes blocks of size 125 m x 125 m; we consider that 1-hop communication can be established between vehicles at either a line-of-sight distance 250 m or non-line-of-sight distance of 140 m. As for highway VANETs, we use samples measured by the Berkeley Highway Laboratory regarding highway I-80 [6], which provides information about the time and speed the vehicles pass through certain points of the highway, which we extrapolate assuming that the vehicles' speeds are constant. From this data, we look at instantaneous snapshots in both urban and highway VANETs, and study the underlying graph where nodes are vehicles and edges exist between vehicles which can establish 1-hop communication. Works in [7] and [8] also study the data from highway I-80 to address the challenge of routing in sparse VANETs. For each scenario, we consider 3 connectivity regimes - high, medium and low connectivity - and the corresponding vehicle densities (we assume a penetration rate of 100%, as well as 100% success rate for the exchanging of messages within the communication ranges, as defined before, and 0% in all other cases). In the case of urban VANETs, this corresponds to inserting more or less vehicles in the CA model, while in the case of highway VANETs, it corresponds to collecting data from different times of the day. This information is summarized in Table I.

TABLE I.     DENSITIES AND RESPECTIVE CONNECTIVITIES FOR URBAN AND HIGHWAY SCENARIOS

| Urban (for 4km$^2$ of area) | | Highway (for 25km of length) | | |
|---|---|---|---|---|
| *Density* | *Connectivity* | *Density* | *Connectivity* | *Time of Day* |
| 10 veh/km$^2$ | ~ 2% | 3.9 veh/km | ~ 3% | 01:00 - 03:00 |
| 60 veh/km$^2$ | ~ 47% | 26.0 veh/km | ~ 68% | 10:00 - 12:00 |
| 80 veh/km$^2$ | ~ 90% | 44.9 veh/km | ~ 98% | 15:00 - 17:00 |

## III. NETWORK PROPERTIES

In this section, we study several parameters of VANETs based on a network graph analysis and in specific cases compare them with well-known theoretical models on node degree distribution, average shortest path length, clustering coefficient, connectivity, and number of clusters. In the following, we define these concepts:

1) **Degree of a node** - the nodes of a network graph are connected to each other through edges that represent links of communication or other types of relationship between them. The number of edges that have a certain node as their start or finish node represents the degree of that node.

2) **Average shortest path length** - this parameter represents the average minimum number of hops between all possible pairs of nodes in a network.

3) **Clustering coefficient** - this coefficient quantifies the clustering of the community of neighbors of a node or the clustering of the network. In the node version, it is defined as the average fraction of pairs of nodes connecting to a certain node that are also connected to each other; in the network version, it represents the probability that two nodes connected to a common node are also connected between themselves.

4) **Connectivity** - this parameter represents the fraction of pairs of nodes in a network which can communicate among themselves. This means that there exists at least one multi-hop path to connect them. For a single node it is defined as the fraction of other nodes in the network this node can communicate with.

We also compare VANETs to small-world and scale-free networks, due to the relevant properties of these networks. A small-world network is a network where the average shortest path length between its nodes is very small compared to the total number of nodes in the network. This is mathematically defined by having an average shortest path length $L$ which grows proportionally to the logarithm of the number of nodes $N$ of the network [1]:

$$L(N) \propto \log(N) \quad (1)$$

Examples of different types of small-world networks can be found in [9][10].

A scale-free network is a special case of a small-world network [11], where the structure of the network remains the same regardless of the scale of the observation [10]. These networks possess many nodes with low degrees and a very small number of nodes with large degrees (and which behave like hubs). A scale-free network shows a node degree distribution that follows a power law (2) at least asymptotically [2]:

$$P_K(k) \propto k^{-\gamma} \quad (2)$$

where $k$ is the degree of the node and $\gamma$ is a constant parameter dependent on the specific distribution.

## A. Node degree distribution

When studying the properties of a network graph, a node's degree is the number of vertices connecting it to other nodes. In this paper, we consider these vertices to appear between any two vehicles within 1-hop distance. Therefore, for VANETs, the node degree distribution represents the probability distribution of the number of vehicles within 1-hop distance from each vehicle.

*1) Urban VANET*

Figure 1 shows the node degree distribution for urban VANETs of different densities in a region of 4km$^2$, as well as the Gaussian curves that best fit these distributions.

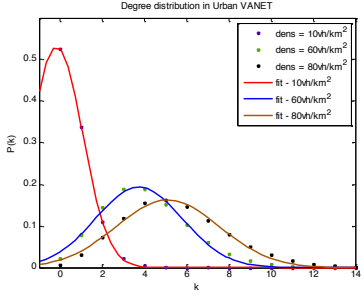

Figure 1. Node degree distributions of urban VANETs with different vehicle densities and their respective Gaussian fits.

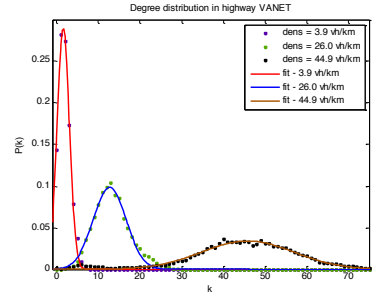

Figure 2. Node degree distributions of highway VANETs with different vehicle densities and their respective Gaussian fits.

These fits are parameterized by the values given in Table II according to the following formula:

$$P(k) = a \cdot e^{-\left(\frac{k-b}{c}\right)^2} \quad (3)$$

where $a, b$ and $c$ are the constant parameters that define these Gaussian distributions. This formula is only valid for $\geq 0$, in contrast with the regular Gaussian distribution.

TABLE II. PARAMETERS OF THE BEST GAUSSIAN FITS FOR THE NODE DEGREE DISTRIBUTIONS IN URBAN VANETs

| Density | a | b | c | R-square | SSE |
|---|---|---|---|---|---|
| 10 veh/km² | 0.5315 | -0.1743 | 1.74 | 0.9999 | 0.00002274 |
| 60 veh/km² | 0.1932 | 3.728 | 2.924 | 0.9941 | 0.000755 |
| 80 veh/km² | 0.1627 | 5.098 | 3.467 | 0.9944 | 0.0005928 |

One can see, from Figure 1 and from the values on Table II, that the Gaussian curves fit very well with the data from the CA model for urban VANETs. One could expect some closeness to scale-free networks due to the possibility that nodes at intersections could work as major hubs to their long ranges of communication in multiple directions. However, since the degree distributions do not follow a power law (2), the networks are not scale-free. Thus, one cannot use the known properties of scale-free networks here, but one can explore the Gaussian model for the node degree distribution to create improved routing protocols, for example.

*2) Highway VANET*

Considering highway VANETs, Figure 2 shows the node degree distribution for different vehicle densities (corresponding to the 3 times of day and traffic patterns) in highway I-80 in Berkeley [6] for a 20km length, as well as the Gaussian curves that best fit these distributions. These fits are parameterized by the values given in Table III according to (3). Once again, the results yield very good fits for a Gaussian model, and show that these VANETs are not scale-free networks. As for urban VANETs, we can explore the Gaussian model for the node degree distribution.

TABLE III. PARAMETERS OF THE BEST GAUSSIAN FITS FOR THE NODE DEGREE DISTRIBUTIONS IN HIGHWAY VANETs

| Density | a | b | c | R-square | SSE |
|---|---|---|---|---|---|
| 3.9 veh/km | 0.2902 | 1.604 | 2.036 | 0.9953 | 0.0009475 |
| 26.0 veh/km | 0.09851 | 12.73 | 5.591 | 0.9904 | 0.000561 |
| 44.9 veh/km | 0.03411 | 45.84 | 15.68 | 0.976 | 0.0003188 |

*B. Average Shortest Path Length*

Another relevant metric is the average shortest path length, which we measure from the real data and fit to known curves.

*1) Urban VANET*

We used the data from the Cellular Automata (CA) model and plotted the average shortest path length as a function of the network area, since this area is proportional to the number of nodes in the network for each density. These fits are parameterized according to (4) with the values of Table IV.

We observe that the power fitness curve fits well for the three types of densities, as shown in Fig 3. Furthermore, for the networks with medium and large densities of nodes, we observe in Table IV that the power exponents (variable 'b') are close to 0.5, which is the exponent of the square root. This shows that, when there is high connectivity, the average shortest path length tends to increase almost proportionally to the radius or to the side of the network.

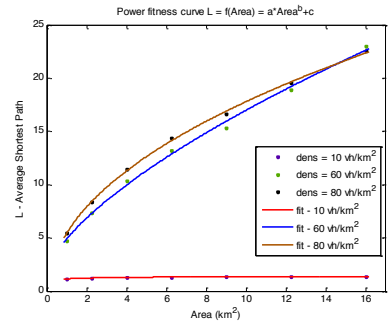

Figure 3. Average shortest path lengths as functions of the network area and vehicle density in an urban scenario, and best fit using the power law

$$L(\text{Area}) = a \cdot \text{Area}^b + c \quad (4)$$

TABLE IV.  PARAMETERS OF THE BEST POWER FITS FOR THE AVERAGE SHORTEST PATH LENGTHS IN URBAN VANETs

| Density | a | b | c | R-square | SSE |
|---|---|---|---|---|---|
| 10 veh/km² | -0.4101 | -0.3173 | 1.557 | 0.9753 | 0.001096 |
| 60 veh/km² | 3.381 | 0.6605 | 1.523 | 0.9963 | 0.9074 |
| 80 veh/km² | 6.505 | 0.462 | -1.044 | 0.9991 | 0.208 |

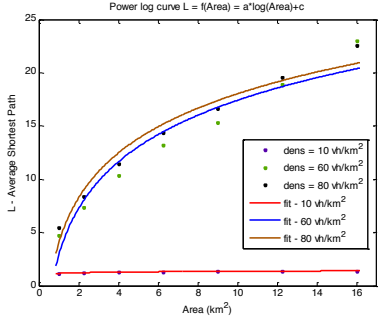

Figure 4. Average shortest path lengths as a function of the network area and vehicle density in an urban scenario, and best logarithmic fits.

Using the logarithmic function (5) we get the curves in Figure 4. Observe that there are significant mismatches for medium and high density of nodes; however, it is a very good fit when the density is low. This is also shown in Table V. This is probably due to the creation of more clusters when the area of the network increases, where these clusters grow slowly on average since there is a low density of nodes. Since for medium and high densities the average shortest path length does not grow with the logarithm of the number of vehicles (which is proportional do the area of the network), these networks are not small-world networks. Lower density urban VANETs have an average shortest path length that does grow approximately with this logarithm, but they have a very low connectivity.

$$L(\text{Area}) = a \cdot \log(\text{Area}) + c \quad (5)$$

TABLE V.  PARAMETERS OF THE BEST LOGARITHMIC FITS FOR THE AVERAGE SHORTEST PATH LENGTHS IN URBAN VANETs

| Density | a | c | R-square | SSE |
|---|---|---|---|---|
| 10 veh/km² | 0.08554 | 1.162 | 0.9594 | 0.001801 |
| 60 veh/km² | 6.312 | 2.889 | 0.9359 | 15.89 |
| 80 veh/km² | 6.073 | 4.052 | 0.9664 | 7.472 |

*2) Highway VANET*

For highway VANETs, a similar study is performed. The obtained results in Table VI for a fit in expression (4) are shown in Figure 5. We observe that the plot showing the power law fits relatively well for the three types of densities, although there is no clear trend for the higher densities, as Figure 5 and Table VI show. The fit using the logarithmic funcion (5) also yields poor results.

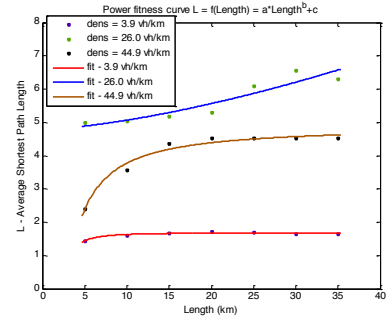

Figure 5. Average shortest path lengths as a function of the highway length and vehicle density in a highway scenario, and best fit using the power law.

TABLE VI.  PARAMETERS OF THE BEST POWER FITS FOR THE AVERAGE SHORTEST PATH LENGTHS IN HIGHWAY VANETs

| Density | a | b | c | R-square | SSE |
|---|---|---|---|---|---|
| 3.9 veh/km | -10.55 | -2.351 | 1.68 | 0.8918 | 0.005471 |
| 26.0 veh/km | 0.008811 | 1.493 | 4.792 | 0.8693 | 0.3402 |
| 44.9 veh/km | -16.8 | -1.184 | 4.873 | 0.9735 | 0.1044 |

*C. Clustering Coefficient*

*1) Urban VANET*

In order to estimate the clustering coefficient of urban VANET's, let us consider a vehicle in an open area which is connected with other vehicles positioned less than $r$ meters away. Assume that this area is infinite and therefore there is no border effect. The clustering coefficient is the probability that 2 vehicles connected to a common vehicle are also connected among themselves: if 2 vehicles are less than $r$ meters away from a common vehicle, clustering coefficient is the probability that these vehicles are also less than $r$ meters away from each other. Assuming the vehicles can be anywhere in the area within $r$ meters of any vehicle with equal probability, we can easily compute the probabilistic distribution of the distance to the common vehicle. The probability of the 2 vehicles being connected is given by the average fraction of the area which is common to the communication range of one of the vehicles and the common vehicle. This probability is the same as the network clustering coefficient $C$ and is given by:

$$C = \int_0^r \frac{2x}{r^2} \cdot \frac{\left[2r^2 \cos^{-1}\left(\frac{x}{2r}\right) - \frac{x}{2}\sqrt{4r^2 - x^2}\right]}{\pi r^2} dx = 0.5865 \quad (6)$$

Comparing the data from the CA model (in Figure 6) with the theoretical values, we observe that that they are similar, as the clustering coefficient is around 0.5 (with 0.5865 being the theoretical value) with no dependence on the network area. For lower densities there seems to be a lower clustering coefficient than for higher densities. These deviations from the theoretical value are due to the border effect resulting from the limited network area, just as the larger confidence intervals should be related to the limited number of vehicles, especially in cases of smaller areas and/or lower vehicle densities.

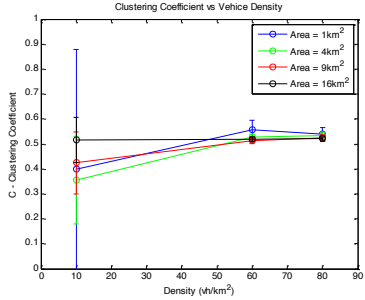

Figure 6. Clustering coefficient as a function of the network area and vehicle density in an urban scenario.

The clustering coefficient provides us with the probability that 2 vehicles which are within the communication range of a common vehicle are also within the communication range of each other. For a value of ~0.5 this means that, if vehicle A has received a specific broadcast message from vehicle B (1-hop communication), then there is a 50% chance that the vehicles within communication range of vehicle A have received the same message from vehicle B (considering a 100% success rate from the layers supporting the message exchange). Thus, if vehicle A would rebroadcast immediately, at least 50% of its neighbors would receive a redundant message (they might even have already received the same message from a source other than vehicles A and B).

*2) Highway VANET*

Through similar derivations for a 1D case, we get the estimate for the clustering coefficient in highway VANETs:

$$C = \int_0^r \frac{1}{r} \cdot \frac{2r-x}{2r} dx = 0.75 \quad (7)$$

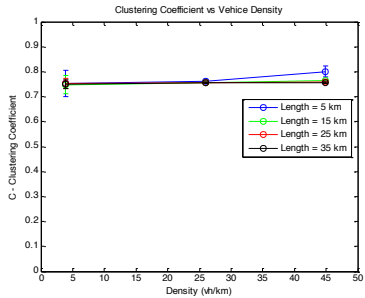

Figure 7. Clustering coefficient as a function of highway length and vehicle density in a highway scenario.

If we compare the values obtained for the clustering coefficient from the I-80 data in Figure 7 and our theoretical value of 0.75, which is not dependent on the highway length, we see that they match almost exactly in most cases, except for the case of the smallest highway length with the highest vehicle density, where the clustering for the I-80 data is slightly higher than that of our model.

From the theoretical model we can observe that there is no expectable dependency on any parameter if we consider the stretch of highway to have infinite length.

*D. Connectivity*

*1) Urban VANET*

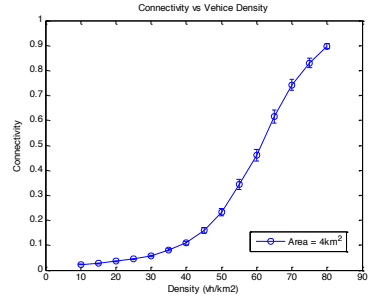

Figure 8. Connectivity as a function of vehicle density for an urban scenario of 2km x 2km

In Figure 8, we observe the connectivity values measured from the data generated by the CA model for urban VANETs. The largest variation of connectivity occurs between 40 and 80 vehicles per $km^2$. Through this figure we can roughly observe the relation between vehicle density and connectivity for urban VANETs.

*2) Highway VANET*

The connectivity values obtained from the expanded data from highway I-80, as a function of the highway length and the vehicle density, are plotted in Figure 9.

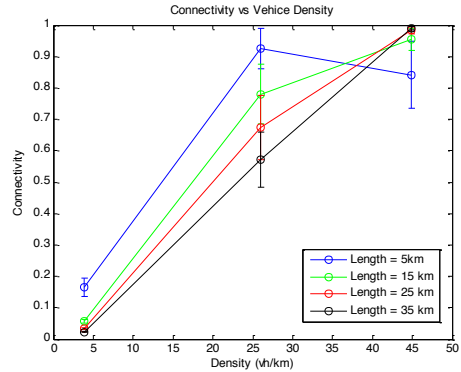

Figure 9. Connectivity as a function of vehicle density and highway length for a highway scenario.

An interesting phenomenon can be observed here: for low and medium vehicle densities, the connectivity is higher for lower highway lengths, while for a high vehicle density, the opposite happens and connectivity increases as the highway length grows. One should note that this length-related difference is very significant, as it can produce a connectivity variation of more than 30% when the road length goes from 5 to 35km. With this data, we can relate vehicle density and connectivity, which will be helpful to design optimized routing protocols and new VANET protocols.

## IV. APPLICATIONS

### A. General Principles

Based on the study of the network properties shown in the previous section, we can infer different ways to improve VANET routing protocols. In particular, based on periodic beaconing, a vehicle can correlate overall vehicle density with its node degree (i.e., node degree indicates the number of its neighbors). This allows a vehicle to determine whether it is the vehicle with the higher number of neighbors in the network, and uses this information to optimize its message relay mechanism. In addition to node degree metric, Section III.C indicates non-zero clustering coefficients for both urban and highway VANETs. This implies that a message broadcast from a vehicle is likely to be redundant with message broadcasts from one of its neighbors. Note that such redundant effect persists regardless of the vehicle density and network size. Due to this observation, it becomes clear that a suppression mechanism could potentially be used to address such issue. As a result, it is interesting to see how the results of the network property analysis presented here can be used to design and/or optimize VANET routing protocols. To illustrate this concept, an urban broadcast routing protocol is used as an illustrative example as shown later in this section.

### B. Improvement of the UV-CAST protocol

Urban Vehicular Broadcast (UV-CAST) [4] protocol is proposed for safety message dissemination applications in urban VANETs, and it aims to deliver safety messages to all vehicles within a specified Region of Interest (ROI). UV-CAST is designed to address both the imminent problems in well-connected and sparse networks (i.e., broadcast storm and disconnected network problems). As shown in Figure 10, a vehicle may operate in one of the two modes: well-connected or disconnected modes. In the well-connected regime, a broadcast suppression technique is implemented: based on its distance from the previous relay and location (e.g., whether it is at the intersection), a vehicle computes the waiting time and only rebroadcast the message if the timer expires and it does not receive a redundant message within the computed wait time. In the case that the vehicle is not in a well-connected regime, it may be assigned the Store-Carry-Forward (SCF) task. Additional details can be found in [4]. While the UV-CAST protocol is shown to yield very good results, it still has some room for improvement as shown in Table VII where the performance of the UV-CAST protocol is compared against the optimal scheme (i.e., a flooding scheme where there are no message collisions or losses). To be specific, the UV-CAST protocol can be further improved in terms of network reachability and average received distance for low vehicle densities scenarios (see Table VII and Figure 11), and in terms of message redundancy for high vehicle densities scenarios (i.e., [4] has shown that a vehicle receives, on average, 3.5 duplicate messages and the number tends to increase with vehicle density (see Figure 11)).

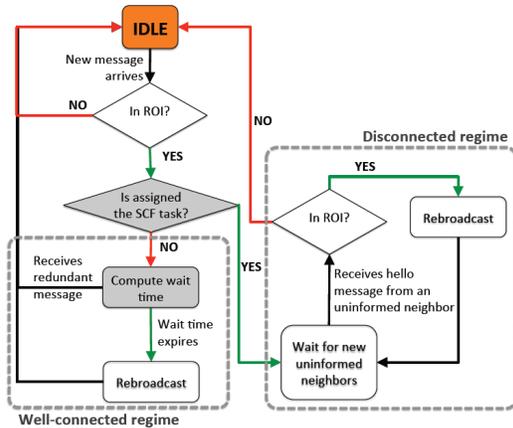

Figure 10. Flowchart describing the operation of the UV-CAST protocol. (Abbrv: ROI-Region of Interest, SCF-Store-carry-forward mechanism) [4].

TABLE VII. AVERAGE NETWORK REACHABILITY OF THE UV-CAST PROTOCOL AND THE OPTIMAL SCHEME IN THE MANHATTAN STREET SCENARIOS [4]

| Network Density [veh/km$^2$] | UV-CAST | Optimal scheme |
|---|---|---|
| 20 | 83.25% | 93.25% |
| 40 | 97.13% | 100% |
| 60 | 99.08% | 100% |
| 80 | 99.63% | 100% |
| 100 | 99.65% | 100% |
| 200 | 99.72% | 100% |
| 400 | 99.89% | 100% |

#### 1) Improving the UV-CAST protocol via network science

As mentioned previously, we aim to improve the UV-CAST protocol and the extension will further address the disconnection and broadcast storm problems in sparse and well-connected regime, respectively. In the proposed improvement techniques, we introduce the variable $k_{med}$, which is defined as the average number of neighbors of a vehicle. Although this variable is calculated based only on local information, it can be used to approximate the overall vehicle density. Note that the results on the node degree distribution shown in Section III.A indicate that it is very unlikely for a vehicle in low density network to have more than 3 neighbors (i.e., $k_{med} < 3$), and most vehicles in high density networks have more than 4 neighbors (i.e., $k_{med} > 4$). Based on such observation, we consider that a vehicle is in a disconnected regime if $k_{med} < 3$ and in a well-connected regime when $k_{med} > 4$. Note that the 'intermediate' regime is not the main focus here, since in such a network (e.g., moderately-connected network), performance of UV-CAST protocol is excellent (i.e., high network reachability and moderate number of redundant messages received at a vehicle). In the proposed scheme, two mechanisms are introduced: one to improve network reachability in low density network, and another to suppress message rebroadcast in high density network.

To improve network reachability in low density networks, we allow vehicles to rebroadcast despite receiving redundant message (before its computed timer expires). While this mechanism may impose additional overhead, it should not be problematic when it is only enabled in low densities scenarios; i.e., despite receiving a redundant message, a vehicle with $k_{med}$ < 3 should still rebroadcast (after the wait time expires) with probability *1-s* where s is defined below:

$$s = \begin{cases} 0.5 + 0.5 \cdot \frac{k_{med}}{3}, & k_{med} < 3 \\ 1, & \text{otherwise} \end{cases} \quad (8)$$

This probability should be smaller as the vehicle density decreases, while maintaining a minimum of 50% probability (since for 50% of the vehicles this message will be redundant on average, according to the average clustering coefficient). We call this mechanism the "s" mechanism.

On the other hand, in order to reduce the number of redundant messages in networks with high vehicle densities, we propose a mechanism to reduce the number of Store-carry-forward (SCF) agents (in the disconnected regime) and number of message rebroadcasts (in the well-connected regime). To be specific, a vehicle only rebroadcasts (after the wait time expires) with probability 'p', and each time a vehicle fulfills the conditions to be an SCF agent, it becomes such an agent with probability $p$.

$$p = \begin{cases} 1, & \text{otherwise} \\ 0.5 + \frac{0.5}{k_{med}-4+1}, & k_{med} > 4 \end{cases} \quad (9)$$

Observe that, the value of *p* decreases with the vehicle density, and it is lower bounded by 0.5 because the average clustering coefficient of 0.5 indicates that half of the times a relayed message will be redundant, and half of the times it will not. We call this the "p" mechanism.

Note that the average shortest path length and clustering coefficient results are not directly utilized in the improvement mechanisms, because both variables require additional message exchanged between vehicles (i.e., 2-hop neighbor information is needed to compute their values).

*2) Tests & Results*

We evaluated the two proposed mechanisms in the simulation environment similar to the one reported in [4]; i.e., a 1km x 1km ROI is considered around an accident scene in a Manhattan grid scenario. In addition, a 15-minute warm-up period is assumed. After the warm-up period, the source vehicle at the center of the network broadcasts a message, and statistics are collected 2 minutes after the first message broadcasts from the source. A total of 10 simulations were run. Note that $k_{med}$ is computed based on an exponential moving average of the number of neighbors of the vehicle.

Figure 11 present simulation results of both p and s mechanisms in terms of four different metrics:

1. Reachability - the fraction of vehicles in ROI that is informed by the end of the simulation;
2. Average received distance - the average distance to the accident scene when the vehicles receive the warning message;
3. Average number of messages received per vehicle;
4. Average number of transmitted messages per vehicle.

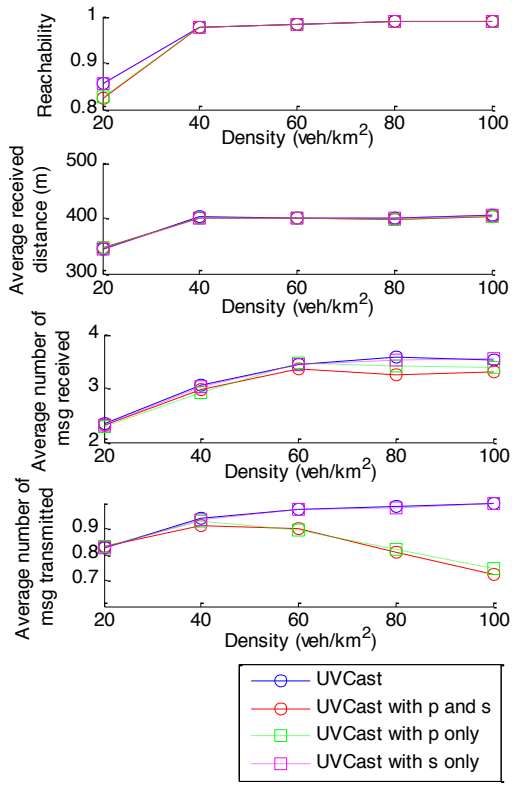

Figure 11. Mechanisms added to the UV-CAST protocol with different combinations of the "p" and "s" mechanism regarding the reachability, the average received distance, the average number of messages received per vehicle and the average number of messages transmitted per vehicle.

In the two plots in the top of Figure 11, the performance of the proposed mechanisms are comparable to that of the UV-CAST protocol in terms of reachability (except for the lowest density) and average received distance. Significant improvement of the proposed mechanisms is, however, observed in the high density networks (see the bottom two plots of Figure 11). It is worth pointing out that the 'p' mechanism can effectively address the broadcast storm problem as it significantly reduces the number of messages transmitted and received at a vehicle; UV-CAST with 'p' mechanism scales well with vehicle density. In addition, observe that the 's' mechanism has negligible effect, especially in the high density networks, because it is only used when the vehicle density is low.

Observing the results for each density, we observe that, for the lowest density considered, there is a slight degradation in reachability from 85% to 82% (see plot for "Reachability") without any change in the overhead on the network (rate of messages transmitted and received; see plots "Average number of msg transmitted" and "Average number of msg received"). Even though this might not be very significant, it goes against the goal of improving the network reachability for low vehicle densities, which is the goal of the designed "s" mechanism. In spite of this, for intermediate and high densities, the improvements are very significant, as we can see that both the "p" and "s" mechanisms contribute (with most contribution from the "p" mechanism) to a significant reduction in the network overhead (see the plots on the average number of transmitted and received messages) without any degradation on the reachability and average received distance metrics (see plots on Reachability and Average Received distance (m)). It is interesting to note that, with the proposed mechanisms (both the "p" and "s" mechanisms), the overhead (in terms of average number of message transmitted per vehicle) decreases with the vehicle density whereas the original UV-CAST provides the opposite trend. For the network with 100 veh/km$^2$ density, ~25% improvement in terms of number of transmitted message per vehicle is observed when the proposed mechanisms (both "p" and "s" mechanisms) are in place.

Thus, it is clear that the proposed "p" mechanism can improve the performance, as it helps in a very significant way to decrease the overhead in the network while maintaining the same performance quality. The "s" mechanism helps the "p" mechanism slightly by reducing the average number of messages transmitted and received. In spite of this, we observe that it does not improve the reachability of the network for the lower densities. It appears that the "s" mechanism acts as a message suppressor: the "s" mechanism enables some messages to be rebroadcasted by vehicles which have received the same message a short time before, but these new rebroadcasts end up suppressing the next rebroadcasts due to their high likelihood of having a redundant character for the multiple receiving vehicles. These effects are barely noticeable for lower densities, but become more visible for higher densities.

While the proposed mechanisms are not effective in low density networks, improvement in terms of overhead in medium and high densities networks seems promising. Additional investigation is therefore needed to completely understand the dynamics of network properties of VANETs, and how one could incorporate network science in designing optimal (or completely new) routing protocols for VANETs.

## V. CONCLUSION

In this work, we have studied urban and highway VANETs from a network science perspective by focusing on parameters such as node degree distribution, clustering coefficient, average shortest path length and connectivity. It is shown that VANETs are not scale-free networks as their node degree distributions can be approximated by Gaussian probability distributions; although low density VANETs might behave somehow similarly to small-world networks, the connectivity is too low to benefit from the small-world property. Furthermore, we established relationships between vehicle density and node degree distribution, as well as connectivity, in order to relate local information and global information. We also showed that the VANETs clustering coefficient is independent from the vehicle density, and we suggested ways to improve existing VANET protocols and/or create new VANET protocols. As an illustrative example, the urban vehicular broadcast protocol, namely UV-CAST protocol, is used as a baseline protocol. Simulation results indicate that the protocol can be improved via network science. The proposed extension to the UV-CAST protocol can significantly reduce the network overhead without degrading the reachability performance especially in medium and high network density scenarios. Nevertheless, it remains an interesting subject for future study to evaluate the impact of factors such as mobility patterns, road topology, penetration rate, and the underlying wireless network on the statistics of key network parameters and routing protocol performance. More importantly, network science can be used as an important tool to determine the performance limits of routing protocols and how one should design VANET routing protocols so that one can achieve such limits.

ACKNOWLEDGMENT

We would like to thank André Cardote for providing us with semi-processed data and useful code for the expansion of the data collected by the Berkeley Highway Lab on the traffic in highway I-80. This work in this paper is supported by the FCT project DRIVE-IN: Distributed Routing and Infotainment through VEhicular Inter-Networking, CMU-PT/NGN/0052/2008.